# Yttrium Tantalum Oxynitride Multiphases as Photoanodes for Water Oxidation

*Wenping Si,[†,‡] Zahra Pourmand Tehrani,[†] Fatima Haydous,[†] Nicola Marzari,[§] Ivano E. Castelli,[//] Daniele Pergolesi,[\*,†,⊥] and Thomas Lippert[\*,†,#]*

[†]Laboratory for Multiscale Materials Experiments, Paul Scherrer Institut, 5232 Villigen PSI, Switzerland. [‡]Key Laboratory of Advanced Ceramics and Machining Technology, Ministry of Education, School of Materials Science and Engineering, Tianjin University, Tianjin 300072, P.R. China. [§]Theory and Simulation of Materials (THEOS) and National Centre for Computational Design and Discovery of Novel Materials (MARVEL), École Polytechnique Fédérale de Lausanne, 1015 Lausanne, Switzerland. [//]Department of Energy Conversion and Storage, Technical University of Denmark, Fysikvej 309, DK-2800 Kgs. Lyngby, Denmark. [⊥]Electrochemistry Laboratory, Paul Scherrer Institut, 5232 Villigen-PSI, Switzerland. [#]Laboratory of Inorganic Chemistry, Department of Chemistry and Applied Biosciences, ETH Zurich, 8093 Zurich, Switzerland.

●, *Supporting Information*

**ABSTRACT:** Perovskite yttrium tantalum oxynitride is theoretically proposed as a promising semiconductor for solar water splitting because of the predicted bandgap and energy positions of band edges. In experiment, however, we show here that depending on processing parameters, yttrium tantalum oxynitrides exist in multiphases, including the desired perovskite $YTaON_2$, defect fluorite $YTa(O,N,□)_4$, and N-doped $YTaO_4$. These multiphases have bandgaps ranging between 2.13 and 2.31 eV, all responsive to visible light. The N-doped YTaO$_4$, perovskite main phase, and fluorite main phase derived from crystalline fergusonite oxide precursors exhibit interesting photoelectrochemical performances for water oxidation, while the defect fluorite derived from low crystallized scheelite-type oxide precursors show negligible activity. Preliminarily measurements show that loading IrO$_x$ cocatalyst on N-doped YTaO$_4$ significantly improves its photoelectrochemical performance encouraging further studies to optimize this new material for solar fuel production.

## ■ INTRODUCTION

Photoelectrochemical (PEC) water splitting driven by solar light offers promises as a clean and effective method to convert solar energy into chemical energy in the form of hydrogen as renewable fuel. The PEC water splitting process includes the absorption of a photon by a semiconductor to create an electron-hole pair, the separation and migration of the two particles to the surface of the semiconductor to oxidize/reduce water into O$_2$ and H$_2$.

To utilize as much as possible solar light, many efforts have been dedicated to develop semiconductor photocatalysts with light absorption extended to the visible spectrum region. Under this perspective, many oxynitrides are promising photocatalyst materials. Compared to the pristine oxides, the bandgaps of many oxynitrides are within the visible light region due to an upward shift in the valence band edge as a consequence of the hybridization of the O 2p and N 2p orbitals.[1-8] In some case, the N substitution also shifts the conduction band edge downward, therefore contributing to the overall reduction of the band gap.[7]

Recent theoretical studies report the result of a computational screening of thousands of possible compositions of perovskite oxynitrides based on the structural stability, the bandgap width, and positions of band edges, which indicates YTaON$_2$ as a promising semiconductor for solar water splitting.[9-10] The structural stability of this compound is however still a controversial issue since other studies predict the perovskite YTaON$_2$ to be structurally unstable as far as the tolerance and octahedral factors are concerned.[11] In literature, only traces of the perovskite structure were experimentally obtained after ammonolysis of the precursor YTaO$_4$, whereas in most cases, defect fluorite YTa(O,N,□)4 (where □ denotes a vacancy) were the main products.[12-16] Although Muraoka et al.[16] labelled the ammonolysis product as pyrochlore Y$_2$Ta$_2$O$_5$N$_2$, the XRD pattern clearly indicates the fluorite type structure with no sur-structure peaks as expected in the case of the pyrochlore phase for larger rare-earth elements.

Inspired by theoretical predictions and based on the available experimental results, we report here a thorough study of the ammonolysis of YTaO$_4$ (Y: Ta = 1 ratio) to revisit and clarify the preparation and PEC performances of the Y-Ta-O-N system for water oxidation.

The literature shows that structure and morphology of the precursor oxide can significantly affect the ammonolysis process.[17] For this reason, in this work, both low crystallized scheelite-type oxide and crystalline fergusonite YTaO$_4$ are used as oxide precursors, prepared by the polymerized complex method and solid state reaction, respectively. Ammonolysis under various temperatures and ammonia flows are used in experiment, resulting in a series of oxynitrides with different crystalline structures and bandgaps. PEC measurements are conducted to examine the photoactivity towards water oxidation of the obtained compounds.

■ **EXPERIMENTAL SECTION**

**Preparation of yttrium tantalum oxynitride**

**Preparation of the oxide precursor YTaO$_4$:** Two methods have been utilized to prepare the oxide precursor in this system. 1) Polymerized complex method (PC): To prepare the oxide precursors, Y(NO$_3$)$_3$·6H$_2$O (Alfa Aesar, 99% Reactor) and TaCl$_5$ (Chempur, 99.99%) were dissolved in methanol (VWR Chemicals) with the addition of citric acid (Sigma, ≥99%) and ethylene glycol (Aldrich, 99.8%) (citric acid and ethylene glycol were added in the solution as concentration ratio of [Y]: [Ta]:[Citric acid]: [ethylene glycol]=1:1:5:20). The solution was stirred at 200 °C to promote polymerization, and eventually a transparent gel was formed with uniformly dispersed constituent elements. Afterwards, the gel was transferred to an alumina crucible and heated at 750 °C for 5 h in air, resulting in low crystallized scheelite-type YTaO$_4$ powders. 2) Solid state reaction method (SSR): Y$_2$O$_3$ (Alfa Aesar, 99.99%, Reactor) powders and Ta$_2$O$_5$ (Alfa Aesar, 99.993%, Puratronic) powders were used as the precursors to prepare crystalline fergusonite YTaO$_4$. Y$_2$O$_3$ and Ta$_2$O$_5$ powders in a stoichiometric ratio of 1:1 were mixed and grinded for 1h, and then pressed into a pellet, which was heated at 900 °C for 2 h and then at 1400 °C for 10 h. In order to get single phase YTaO$_4$, this procedure was repeated four times.

**Preparation of yttrium tantalum oxynitride through thermal ammonolysis:** 0.15 g of the oxide precursor were put in an alumina boat and placed in a tube furnace through which ammonia gas flowed at a rate of 50 mL min$^{-1}$ or 220 mL min$^{-1}$. Various temperatures between 850 °C and 950 °C have been used to prepare different structures. NaCl, KCl and NH$_4$Cl were also tried as a mineralizer for the ammonolysis.

**Materials characterizations**

X-ray diffraction (XRD) characterizations of the yttrium tantalum oxides and oxynitrides were conducted using a Bruker−Siemens D500 X-ray powder diffractometer equipped with Cu Kα radiation (1.5418 Å) and a scintillation counter detector. θ-2θ scans were performed with a step increment of 0.01 ° at a speed of 1 s per step. A Cary 500 Scan UV−Vis−NIR spectrophotometer equipped with an integrating sphere was used to measure the diffuse reflectance of the powders to obtain the band gaps. The measured reflectance was converted to Kubelka–Munk function, and the band gap magnitudes were estimated by extrapolating the absorption edge to the ground level. Scanning electron microscopy (SEM) images were obtained with a Zeiss Supra VP55 Scanning Electron Microscope. The N content of the oxynitride powders was determined by thermogravimetric (TG) analysis from the mass difference between the oxynitride and the resulting oxide after heating in synthetic air. The TG measurements were acquired using a NETZSCH STA 449C analyzer equipped with PFEIFFER VACUUM ThermoStar mass spectrometer, where oxynitride powders were heated in alumina crucibles to 1400 °C with a heating rate of 10 °C min$^{-1}$ in 36.8 mL min$^{-1}$ synthetic air.

**Photoanode preparation**

Yttrium tantalum oxynitride photoanodes were prepared *via* the electrophoretic deposition method (EPD) followed by a necking treatment.[18] In detail, 120 mg yttrium tantalum oxynitride powders were dispersed in 50 mL acetone with 10 mg iodine and sonicated for 60 min. The yttrium tantalum oxynitride powders were positively charged due to the releasing of H$^+$ when iodine reacted with acetone. Two parallel FTO substrates (~ 7 Ω sq$^{-1}$, Sigma Aldrich) were then placed in the oxynitride dispersion with a distance of 1 cm under a bias of 20 V for 5 min. Yttrium tantalum oxynitride powders were deposited on the negative electrode. The so-called necking treatment was then applied to the electrode to improve the electronic conduction between oxynitride particles and the FTO substrate. Yttrium tantalum oxynitride photoanodes were treated with 25 μL of 10 mM TaCl$_5$ methanol solution and dried in air. This procedure was repeated for three times, followed by heating at 350 °C in air for 30 min. This temperature was chosen to avoid oxidation of the oxynitrides.

Cocatalyst IrO$_x$ was post-loaded on the oxynitride photoanodes by impregnation from a colloidal IrO$_2$ aqueous solution, which was prepared by hydrolysis of Na$_2$IrCl$_6$.[19] The colloidal IrO$_2$ was prepared as follows: 50 mL H$_2$O was used to dissolve 0.008 g Na$_2$IrCl$_6$·6H$_2$O, the pH of which was adjusted to 11–12 with 1 M NaOH. The solution was heated at 80 °C for 30 min and cooled to room temperature by immersion in an ice water bath. HNO$_3$ was used to adjust the pH of the cooled solution to 10. A blue solution containing colloidal IrO$_2$ was obtained by subsequent heating at 80 °C for 30 min. The photoanodes were immersed in the IrO$_2$ colloid for 30 min, heated at 300 °C for 30 min and then washed with distilled water before PEC measurements.

**PEC measurements**

PEC measurements were performed in a three electrode configuration in 0.5 M NaOH (pH=13.0) aqueous solution. The yttrium tantalum oxynitride photoelectrode was used as the working electrode; a coiled Pt wire and Ag/AgCl were used as the counter and reference electrodes, respectively. To simplify the display of the working electrode potential, V$_{RHE}$ (potential *vs*. reversible hydrogen electrode) was adopted throughout the manuscript, which was obtained according to the Nernstian relation. A Solartron 1286 electrochemical interface was used to carry out the voltage scan and current collection. Potentiodynamic measurements with a scan rate of 10 mV s$^{-1}$ in the potential window of 0.5−1.7 V$_{RHE}$ were performed to investigate the PEC performance of Y-Ta-O-N multiphases. The chopped dark−light current densities were normalized according to the illuminated area. The light source was a 150 W Xe lamp equipped with an AM 1.5G filter (100 mW cm$^{-2}$, Newport 66477-150XF-R1) calibrated by a photodetector (Gentec-EO).

■ **RESULTS AND DISCUSSION**

**Structural analysis of Y-Ta-O-N multiphases**

As shown in Figure 1 and consistent with literature,[12] oxide precursors obtained by polymerized complex method were indexed to the XRD pattern of low crystallized scheelite-type YTaO$_4$ (denoted as LO, PDF #00-050-0846). Various temperatures and the flow rates of ammonia were tested to prepare the oxynitrides. At a high temperature of 950 ºC and under a high ammonia flow of 220 mL min$^{-1}$, yellowish and reddish orange products were obtained after 10 h and 20 h ammonolysis, respectively. The XRD pattern shown in Figure 1 indicates a defect fluorite structure YTa(O,N,□)$_4$, as reported in literature.[12] In the following, we refer to these samples as "fluorite-nitride" FN1 and FN2.

In agreement with the available literature,[12] we can confirm that low crystallized scheelite-type YTaO$_4$ tends to be nitrided into defect fluorite YTa(O,N,□)$_4$ at temperatures higher than 900 ºC. However, no perovskite structure can be obtained using the low crystallized scheelite-type YTaO$_4$ precursor.

Solid state reaction method was also used to prepare crystalline fergusonite YTaO$_4$ (denoted as crystalline oxide CO) as precursor for thermal ammonolysis. The XRD pattern in Figure 2 shows that the obtained YTaO$_4$ is in fergusonite crystalline structure (PDF #01-072-2018). After 15 h ammonolysis at 850 ºC, a yellow product was obtained. The XRD pattern indicates that the crystalline structure of the product remains similar to that of the crystalline oxide precursor, with all diffraction peaks only slightly broadened and shifted towards smaller 2θ angles due to N doping in the lattice. We label this product as N-doped YTaO$_4$ "NO".

In addition, a new peak appears at 2θ of ~31.9 °, indicated by the blue arrow in Figure 2, which is similar to the characteristic peak at ~31.3 ° of the perovskite LaTaON$_2$ (PDF #00-053-0960). And the larger 2θ value of this peak for YTaON$_2$ is also consistent with the fact that Y has a smaller atomic radius compared to La. This suggests that the nitridation onset temperature is around 850 °C. But at this temperature, the reaction is so slow that it takes 100 h of ammonolysis to completely convert the oxides to oxynitrides. The same products can be achieved by ammonolysis at 880 °C for 45 h, as shown in Figure 2. In both cases, the products exhibit a yellowish orange color. The XRD patterns confirm that the ammonolysis under 850 °C and 880 °C result in similar products which are a mixture of perovskite YTaON$_2$ (main phase), defect fluorite YTa(O,N,□)$_4$ and a small trace of Ta$_3$N$_5$. These two products are denoted as PM1 and PM2, where PM stands for "perovskite main phase".

Increasing the ammonolysis temperature to 950 °C also leads to a mixture of perovskite and fluorite oxynitrides, but the XRD pattern suggests that the fluorite is the main phase. We refer to this product as "fluorite main phase" FM. Traces of Ta$_3$N$_5$ are also detected by XRD.

Summarizing the results shown in Figure 1 and 2, we conclude that low crystallized scheelite-type YTaO$_4$ tends to be nitrided into defect fluorite oxynitride. Crystalline fergusonite YTaO$_4$ tends to form a mixture of perovskite and defect fluorite where the perovskite is the main phase at temperatures lower than 900 °C and the fluorite is the main phase at higher temperatures. This is most probably due to the phase transition between fergusonite and scheelite YTaO$_4$ upon heating.[20-23] In fact, three crystal structures exist for YTaO$_4$: scheelite (tetragonal, T), fergusonite (monoclinic, M), and a third monoclinic fergusonite form M′, which is very similar to M except for a halved *b* axis.[20-23] T can be obtained at temperature higher than 1450 °C, which transforms to M by cooling, while M′ structure is obtained when crystals are grown below the M-T transformation temperature.[20] Upon heating, the M′ phase transforms sluggishly into T.[20-23]. In this work, 1400 °C was used to prepare YTaO$_4$, which actually leads to M′. Thus, upon heating, the M′ phase transforms sluggishly into T. At lower ammonolysis temperatures, the transformation from fergusonite YTaO$_4$ into the scheelite structure is very sluggish and fergusonite is the main phase, resulting in more perovskite main phase in the ammonolysis products. Instead, at an ammonolysis temperature of 950 °C, the transformation of fergusonite YTaO$_4$ into the scheelite structure becomes faster, leading to more defect fluorite phase in the final oxynitrides.

Chloride salts were examined as mineralizers (also known as flux)[24-26] in the thermal ammonolysis process. It was found that NaCl and KCl promote the decomposition of binary metal oxides eventually resulting in large amounts of Ta$_3$N$_5$.[24] Figure S1 in the supporting information shows the results obtained with NaCl. NH$_4$Cl was also tested as mineralizer but no significant effect was observed. Table 1 shows the summary of the preparation for yttrium tantalum oxides and oxynitrides.

In addition, TG measurements were performed to determine the N contents of the oxynitrides and the results are shown in Figure S2. According to these measurements, the composition of single phase defect fluorite FN2 can be calculated to be YTaO$_{2.47}$N$_{1.01}$□$_{0.52}$, which is very close to the compositions of YTaO$_{2.76}$N$_{0.83}$□$_{0.41}$ and YTaO$_{2.39}$N$_{1.08}$□$_{0.53}$ for defect fluorite phase reported by Maillard et al.[12] Regarding the samples with main phase NO, FM, and PM1, only the nominal compositions can be considered since these samples are not composed by single phases. The nominal compositions of NO, FM and PM1 are YTaO$_{3.32}$N$_{0.45}$, YTaO$_{2.18}$N$_{1.21}$ and YTaO$_{3.01}$N$_{0.66}$, respectively.

**Morphologies of Y-Ta-O-N multiphases**

SEM images in Figure 3a and b show that FN1, FN2 (i.e. obtained from low crystallized scheelite-type LO) have similar morphologies, consisting of nanoparticles with average sizes of a few tens nm. These nanoparticles agglomerate to form large loosely connected aggregates.

The SEM micrographs of NO, PM1 and FM produced from crystalline precursor CO are shown in Figure 3c, d, and e. They exhibit similar morphologies, but in this case micro-powders with grain sizes in the range of 0.5-3 μm are obtained. It is also noteworthy that the grains of micro-powders show large amounts of cracks and pores, probably formed during the high temperature ammonolysis.

These measurements confirm that the oxide precursors have crucial influence on the final oxynitrides. Similar observations have been reported on other oxynitrides such as $LaTiO_2N$,[27] $LaTaON_2$,[28-29] and $CaNbO_2N$.[4]

**UV-Vis absorbance and bandgap**

It can be observed visually that the prepared oxynitrides exhibit distinct colors, as listed in Table 1, indicating different bandgaps ($E_g$). $E_g$ is a very important parameter for light absorption and further photoelectrochemical performances. Diffuse-reflectance absorption spectroscopy was used to estimate $Eg$ for the the fluorite FN1 and FN2, the N-doped $YTaO_4$ NO, the perovskite main phase PM1 and the fluorite main phase FM.

Figure 4 shows the Kubelka-Munk absorbance derived from UV-Vis diffuse reflectance spectrum. FN1 and FN2 obtained from LO exhibit higher absorbance than NO and PM1 obtained from CO. The bandgaps for FN1, FN2, NO, PM1, and FM are 2.25, 2.13, 2.31, 2.23, and 2.26 eV. Considering the correlation of the bandgaps with the samples' thermal history, it is found that the bandgap tends to decrease with higher ammonolysis temperature and longer reaction time. This is most probably due to a higher N content. Based on the bandgaps, a preliminary DFT calculation was also carried out to search for possible structures and the results are shown in Figure S3.

**PEC performance of yttrium tantalum oxynitrides for water oxidation**

PEC measurements for water splitting under chopped AM 1.5 G illumination were performed on photoelectrodes prepared by depositing the oxynitride powders on FTO substrates, and the photocurrent densities are shown in Figure 5a. The oxynitrides obtained from low crystallized scheelite-type oxide precursors, i.e. FN1, and FN2, show negligible photoactivity. While NO, PM1, and FM show clearly measurable photoactivity, with photocurrent densities for water oxidation of about 1.0 µA cm$^{-2}$, 1.9 µA cm$^{-2}$, and 0.6 µA cm$^{-2}$ at 1.3 V$_{RHE}$, respectively. Such a different behavior could be ascribed to the different morphologies observed for the two sets of powders, as shown in Figure 3. The much more extended grain boundary region in the nano-powders obtained from low crystallized scheelite-type oxide precursors increases the recombination of photogenerated charge carriers. Muraoka et al.[16] reported evident photocurrents for yttrium tantalum oxynitride obtained from precursors prepared using the PC route. However, the light source for the photoelectrochemical measurement in this reference was different from what is used in the current work. The power density of the irradiated light in the reference was not calibrated, only mentioning the use of a xenon lamp (300 W) with cut-off filters for visible light irradiation (λ > 500 nm). Depending on the distance of the light source and electrodes, a 300 W xenon lamp could give very different power densities, which is directly related with the measured photocurrent density.

To further improve the photocurrent, cocatalysts are used to suppress the surface recombinations of charge carriers and to promote the water oxidation reaction.[30] In this study, IrO$_x$ cocatalysts are loaded onto the photoelectrodes and the photocurrents are shown in Figure 5b. The product NO shows a photocurrent density of ~17 µA cm$^{-2}$ at 1.3 V$_{RHE}$, while with PM1 and FM photocurrent densities of ~3.2 µA cm$^{-2}$ and ~2.5 µA cm$^{-2}$ at 1.3 V$_{RHE}$ were measured. These results indicate that, compared to the mixed phases of PM and FM, NO has better bulk properties in terms of charge separation and conductivity.

The photocurrent density of IrO$_x$ loaded-NO is comparable to those reported in literature for other well-known oxynitrides such as BaTaO$_2$N,[31] LaTaON$_2$[6, 29], CaNbO$_2$N[4] and Na$_x$La$_{1-x}$TaO$_{1+2x}$N$_{2-2x}$.[32] The PEC performance of this material could be further improved by optimization of the N content to reduce the bandgap and to enhance the conductivity and by loading more suitable cocatalysts to minimize the loss of surface charge carriers.[31] N-doped YTaO$_4$ is indeed a new material that certainly deserves further research studies to fully investigate its potentials for solar water splitting.

■ **CONCLUSIONS**

In this study, different phases of yttrium tantalum oxynitride are synthesized using low crystallized scheelite-type YTaO$_4$ and crystalline fergusonite YTaO$_4$ as the starting oxide precursors. It is found that low crystallized scheelite-type YTaO$_4$ precursor tends to be nitrided into defect fluorite YTa(O,N,□)$_4$ nanoparticles. Crystalline fergusonite YTaO$_4$ precursor tends to be nitrided into micro-powders with mixed crystalline structures including perovskite YTaON$_2$, defect fluorite YTa(O,N,□)$_4$ and a small trace of Ta$_3$N$_5$. Perovskite exists as the main phase in the product of ammonolysis below 900 ºC, while defect fluorite is the main phase above that temperature. In addition, other multiphases can also be obtained such as N-doped YTaO$_4$, which can be synthesized starting from the crystalline oxide precursor through a short ammonolysis of 15 h at relatively low temperature of 850 °C. UV-Vis absorbance measurements show that the bandgaps of these multiphases range between 2.13-2.31 eV.

Although oxynitrides obtained from low crystallized scheelite-type YTaO$_4$ have better visible light absorption properties, these materials do not show any significant photoactivity towards water oxidation. On the contrary, the N-doped YTaO$_4$, perovskite main phase, and fluorite main phase derived from crystalline fergusonite YTaO$_4$ exhibit interesting photoelectrochemical performances. Loading IrO$_x$ cocatalyst improves the photocurrent of N-doped YTaO$_4$ by almost 20 times, enabling the performance to be comparable to other well-known oxynitrides. This work identifies N-doped YTaO$_4$ as a new material showing promising potentials for solar water splitting.

## ■ ACKNOWLEDGEMENTS

This research is supported by the Paul Scherrer Institute and the NCCR MARVEL, funded by the Swiss National Science Foundation.

## ■ ASSOCIATED CONTENT

●, **Supporting Information**

Electronic Supporting Information (ESI) available: XRD patterns of ammonolysis products when using NaCl; TG results; DFT calculation of band gaps. See DOI: 10.1039/x0xx00000x

## ■ AUTHOR INFORMATION

**Corresponding Author**

Contact information for the author(s) to whom correspondence should be addressed.

thomas.lippert@psi.ch; daniele.pergolesi @psi.ch

**Present Addresses**

† present address for F. Haydous is Division of Applied Physical Chemistry, Department of Chemistry, KTH – Royal Institute of Technology, 11428 Stockholm, Sweden.

■ **REFERENCES**

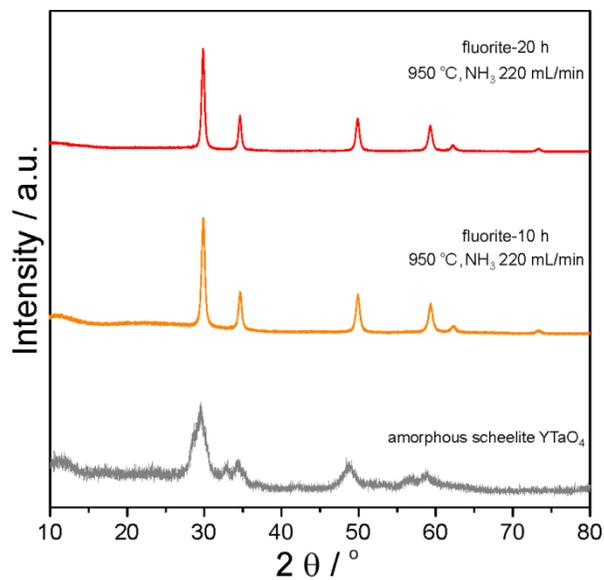

**Figure 1.** XRD patterns of low crystallized scheelite-type YTaO$_4$ and the oxynitride products under various ammonolysis conditions.

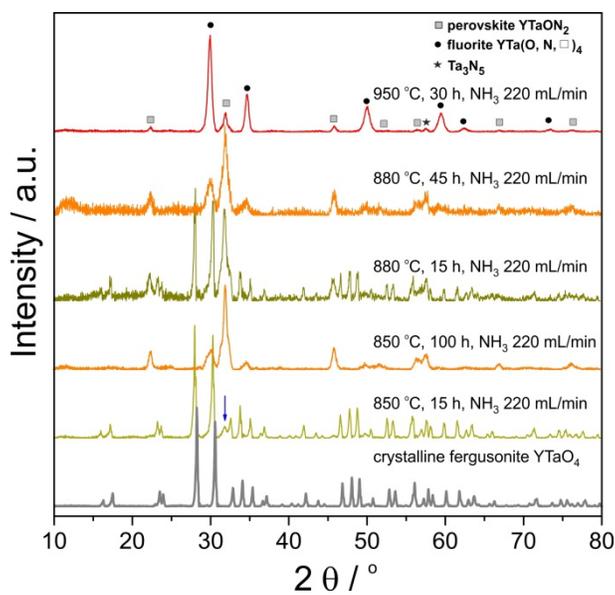

**Figure 2.** XRD patterns of crystalline fergusonite YTaO$_4$ and the oxynitride products under various ammonolysis conditions.

**Table 1.** Preparation conditions for yttrium tantalum oxides and oxynitrides

| XRD phase | denotation | color | starting material | heating treatment |
|---|---|---|---|---|
| Low crystallized scheelite-type $YTaO_4$ | LO | White | $Y(NO_3)_3 \cdot 6H_2O$, $TaCl_5$ | 750 °C, air, 5 h |
| Crystalline fergusonite $YTaO_4$ | CO | White | $Y_2O_3$, $Ta_2O_3$ | 1400 °C, air, 10 h |
| Fluorite $YTa(O,N,\square)_4$ | FN1 | Yellowish orange | LO | 950 °C, $NH_3$ 220 mL min$^{-1}$, 10 h |
| Fluorite $YTa(O,N,\square)_4$ | FN2 | Reddish orange | LO | 950 °C, $NH_3$ 220 mL min$^{-1}$, 20 h |
| N-doped $YTaO_4$ | NO | Yellow | CO | 850 °C, $NH_3$ 220 mL min$^{-1}$, 15 h |
| Perovskite (main phase), fluorite and a small trace of $Ta_3N_5$ | PM1 | Yellowish orange | CO | 850 °C, $NH_3$ 220 mL min$^{-1}$, 100 h |
| Perovskite (main phase), fluorite and a small trace of $Ta_3N_5$ | PM2 | Yellowish orange | CO | 880 °C, $NH_3$ 220 mL min$^{-1}$, 45 h |
| Perovskite, fluorite (main phase) and a small trace of $Ta_3N_5$ | FM | Reddish orange | CO | 950 °C, $NH_3$ 220 mL min$^{-1}$, 30 h |

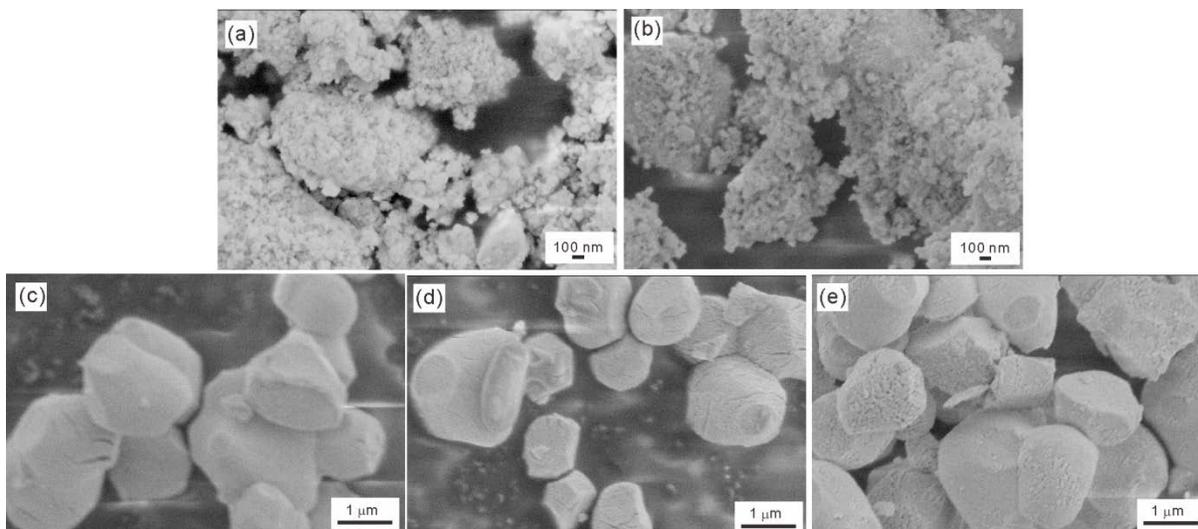

**Figure 3.** SEM images of Y-Ta-O-N multiphases: (a-b) obtained from precursor LO: FN1, FN2. (c-e) from precursor CO: NO, PM1, FM.

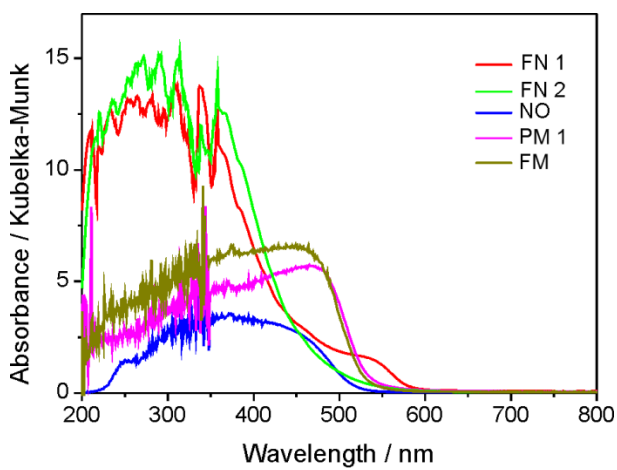

**Figure 4.** Kubelka-Munk absorbance of Y-Ta-O-N with different structures.

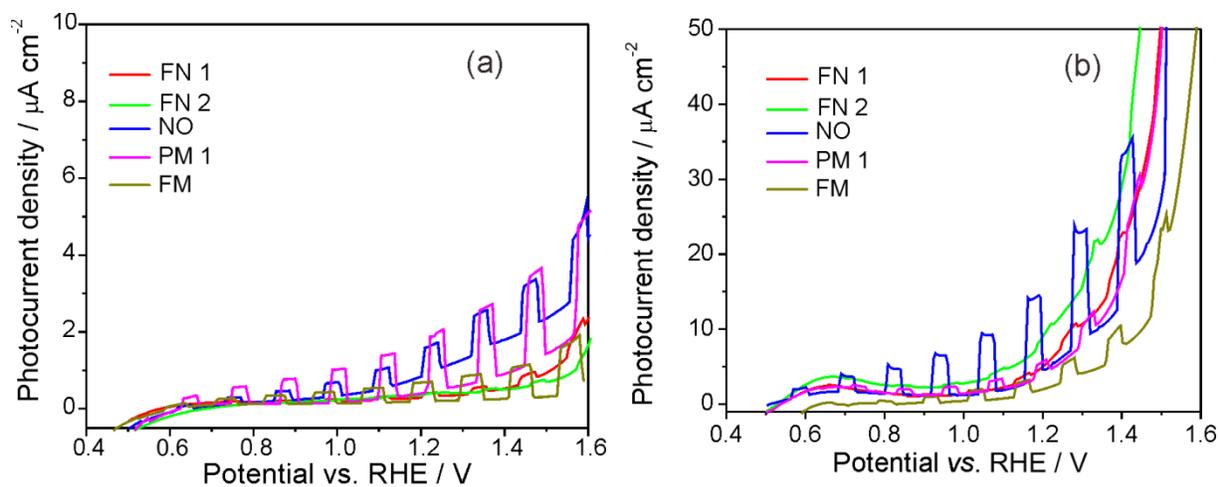

**Figure 5.** PEC performances in 0.5 M NaOH electrolyte (pH=13.0) under AM 1.5G illumination. (a) Y-Ta-O-N powders with different structures, and (b) after loading of $IrO_x$ cocatalysts.